\documentclass[a4paper]{scrartcl}

\usepackage[utf8]{inputenc}

\usepackage{graphicx}
\usepackage{amsmath}
\usepackage{amssymb}
\usepackage{amsbsy}
\usepackage{xspace}

\usepackage[]{algorithm2e}
\usepackage{color}
\usepackage{tabularx}
\usepackage{subfig}
\graphicspath{{fig/}{./}}

\usepackage{hyperref}

\newcommand{\codeurl}{\url{https://gitlab.mpcdf.mpg.de/mpcdf/hpcmd}}
\newcommand{\docurl}{\url{http://mpcdf.pages.mpcdf.de/hpcmd}}

\def\hpcmd{\emph{hpcmd}\xspace}
\def\splunk{Splunk\xspace}

\begin{document}

\title{MPCDF HPC Performance Monitoring System: Enabling Insight via
Job-Specific Analysis}

\author{
Luka Stanisic,
Klaus Reuter
\\
\\
Max Planck Computing and Data Facility\\
Gießenbachstraße 2, Garching, 85748, Germany
\\
\\
\texttt{\{luka.stanisic, klaus.reuter\}@mpcdf.mpg.de}
}


\maketitle

\begin{abstract}
This paper reports on the design and implementation of the HPC
performance monitoring system deployed to continuously monitor
performance metrics of all jobs on the HPC systems at the Max Planck
Computing and Data Facility (MPCDF). Thereby it reveals important
information to various stakeholders, in particular to users,
application support, system administrators, and management.
On each compute node, hardware and software performance monitoring
data is collected by our newly developed 
lightweight open-source \hpcmd middleware
which builds upon standard Linux tools. The data is transported via
rsyslog, and aggregated and processed by a \splunk system, enabling
detailed per-cluster and per-job interactive analysis in a
web browser. Additionally, performance reports are provided to the
users as PDF files. Finally, we report on practical experience and
benefits from large-scale deployments on MPCDF HPC systems,
demonstrating how our solution can be useful to any HPC center.
\end{abstract}

\section{Introduction}
\label{sec:intro}

HPC systems are highly expensive facilities that are rapidly evolving
with respect to computational power, complexity, and size. More and more
scientific disciplines use HPC resources in their research
process to gain insight from numerical simulations or from data
analytics.
Hence, it is essential to strive to
maximize the performance of the applications running on these precious
resources.
However, an efficient usage requires expert knowledge in parallel algorithms and programming,
and a lot of effort spent on optimization and parameter tuning.
This point became more important in recent years with the
advent of processors with many cores and accelerators, which made parallel
programming even more complex.
Having performance numbers available for each job is therefore
essential for the stakeholders of the HPC system, first, to make them aware of potentially
suboptimal usage of resources, and second, to enable them to take action
to improve the way these resources are used.

Jobs on a HPC cluster are commonly orchestrated by a batch scheduler,
which can easily provide usage statistics based
on \emph{allocated} resources. These are often quantified in terms of
CPU or GPU hours, and have proven useful for accounting purposes.
However, these numbers do not carry information about the actual
resource \emph{utilization}.
Performance metrics measured for each job are therefore crucial to learn,
e.g., about under-utilization of allocated resources (idle vector units, or
idle cores and accelerators), or other problematic usage patterns.

Modern hardware provides a plethora of counters that can be
used for performance monitoring. In addition to the arithmetic units, CPUs
have performance monitoring units (PMUs) that can be programmed to count
certain instructions (e.g., scalar and vectorized floating point operations)
with very little performance overhead. Hardware such as GPUs and network
adapters provides similar counters.
These hardware-related metrics can be complemented by software-related metrics,
obtainable from the Linux kernel or from system tools. Such metrics include
information on the running processes, their memory footprint, filesystem-related
counters, etc.

Selecting and efficiently collecting these metrics is a challenge which we
address in the present work. We developed a new lightweight software daemon,
\hpcmd\footnote{hpcmd stands for HPC monitoring daemon.}, that runs on each node,
performs measurements periodically in the
background, and finally writes the data to the syslog.
The syslog lines from all nodes are then propagated to the \splunk framework,
for which we have developed special dashboards to perform advanced interactive data analysis.
As a service to the users, we also provide PDF reports downloadable for
each job.
These two main components, \hpcmd and \splunk dashboards together with few
additional scripts compose a comprehensive suite designed to continuously
monitor the performance of all jobs on the HPC systems at the MPCDF.
We believe that other centers could also benefit from our system.

In the following, we first elaborate on the insight and benefits the various
stakeholders of an HPC system may draw from a performance monitoring system.
Second, we discuss related work
before we describe in detail our solution in the main part of the paper.
Finally, we illustrate several cases in which our system has already proven
very useful, before closing with a summary.

\section{Benefits from an HPC Performance Monitoring System}
\label{sec:benefits}

The following four groups of key stakeholders of an HPC system benefit
from the insight enabled by HPC performance monitoring data.

\paragraph{Computational scientists} and other users who run jobs on an
HPC system typically have to apply for CPU hours.
They have a strong intrinsic motivation to use the
resources as efficiently as possible, in order to maximize the scientific
knowledge they can obtain from the results. Based on HPC monitoring data,
experienced users are often capable of identifying and fixing issues
themselves, e.g., by applying appropriate compiler optimization for a
specific architecture. Less skilled users might be
motivated to approach application support when facing poor performance
indicated by monitoring data.

\paragraph{Application support} at a computing center
provides technical support and is in charge of porting and optimizing
applications for the HPC systems.
HPC performance monitoring data enables application support to detect
problematic jobs, and consequently, to proactively approach users
who are potentially in need of assistance.

\paragraph{System administrators} may benefit from
performance monitoring data, e.g., to better judge the impact of software
updates, security patches, and hardware settings. Potential changes in
application performance after some maintenance work can be traced in an
objective way based on current and historical performance data.

\paragraph{Management} is interested in learning performance numbers that
represent the actual resource utilization in addition to knowing the
allocation of plain CPU hours, a metric that has been widely used up to now
to quantify the resource share.
Moreover, performance data gathered on present systems can be used to steer
decisions for the procurement of future HPC clusters. For example, looking at a
roofline plot with measurement data from most used applications enables
decision makers to judge quickly if these applications are limited by the memory
bandwidth or by the peak floating-point performance, and thereby if investing in new
architectures with higher memory bandwidths would pay off. Similarly,
analysis of network traffic may hint at
applications that would benefit, e.g., from higher network bandwidth or lower latency.
Finally, performance data documents to which degree GPUs are actually
used, especially on multi-GPU nodes.
In these respects, HPC performance monitoring data helps to close
important information gaps.

\section{Related Work}
\label{sec:rl}

There are at least two big challenges regarding the implementation of a HPC
performance monitoring system that have been addressed by various solutions
in recent years.

The first, data-related, challenge is to choose which metrics should
be tracked, how to interpret the collected data, how to identify
performance bottlenecks, and how to ultimately detect if there is a
significant problem in an application code.
There are several software tools that can be used to analyze the performance
of a running job. For example, for CPU codes, there are Linux
perf\cite{PERF2013}, PAPI\cite{PAPI1999}, LIKWID\cite{LIKWID2010}, and
VTune\cite{VTUNE2008}, among others.
These tools provide access to hardware counters which are then often analyzed
using a ``top-down'' method\cite{TOPDOWN2014}. One compares the counter
values to the theoretical peak values of the machine and deduces how well the
compute resources are utilized.
However, there are cases when utilization values appear rather low even
for well-optimized applications, e.g., due to the nature of the
problem the code is solving or the required data structures. Hence,
looking only at the utilization numbers can be misleading, and one
needs to be careful before declaring that a job has a performance
issue. To alleviate this effect, some researchers prefer to rely on
cross-comparisons between different runs and applications
\cite{TACC2014}, recently proposing machine learning
techniques for such analysis\cite{OAKRIDGE2017,ML2019}.

The second, technical, challenge is to design and deploy a system that
works reliably on (multiple) large HPC clusters while introducing minimal
overhead, efficiently collects the data from many nodes into a centralized
database, and provides a powerful framework for analysis and visualization.
For example, the \emph{TACC stat} framework has been developed to achieve these
goals\cite{TACC2014,TACC2016}.
It combines information collected by various standard Linux tools and some custom tools,
e.g., \emph{REMORA}\cite{REMORA2015},
to monitor resource utilization at the Texas Advanced Computing Center.
Next, the \emph{PerSyst} monitoring system developed at the Leibnitz
Supercomputing Center comprises a hierarchical system of collectors
and aggregators, a central database and a web interface to monitor
large-scale HPC systems \cite{LRZ2014}. Thanks to the data aggregation
using quantiles, this tool is well suited for jobs that run on a large
number of nodes.
The \emph{LIKWID Monitoring Stack} targets small to medium
scale systems \cite{LIKWID2017}. It is partly based on the LIKWID
performance tool suite developed by the same group of
authors \cite{LIKWID2010}.
Finally, the \emph{Lightweight Distributed Metric Service (LDMS)} was developed for
performance monitoring at Sandia National Labs \cite{BLUEWATERS2014}.
This framework provided very useful information for the system administrators
and users, while having minimal impact on the application performance.

All the aforementioned solutions gave us valuable ideas and helped
us to better define the goals for our approach. However, there are
several reasons why we decided to develop our own system.
Most importantly, all of these systems either rely on data-measurement
software or on infrastructure setups (e.g., batch system configuration) that are specific to the
center where they have been developed, and hence, would be difficult to adapt and maintain.
Moreover, many of the existing approaches appear to rely on complex
hierarchical communication layers and custom web-based visualization
platforms, while we found ourselves in the convenient position to use rsyslog
and \splunk systems that had already been deployed at the MPCDF for other
monitoring purposes, e.g., the monitoring of the system ``health'' status.

\section{Solution Architecture}
\label{sec:solution}

In this section, we detail on how our system can obtain, collect, analyze, and
present performance data from HPC clusters, addressing the needs of
all stakeholder groups mentioned in Section \ref{sec:benefits}.

\begin{figure}[tb]
\centering
\includegraphics[width=\textwidth]{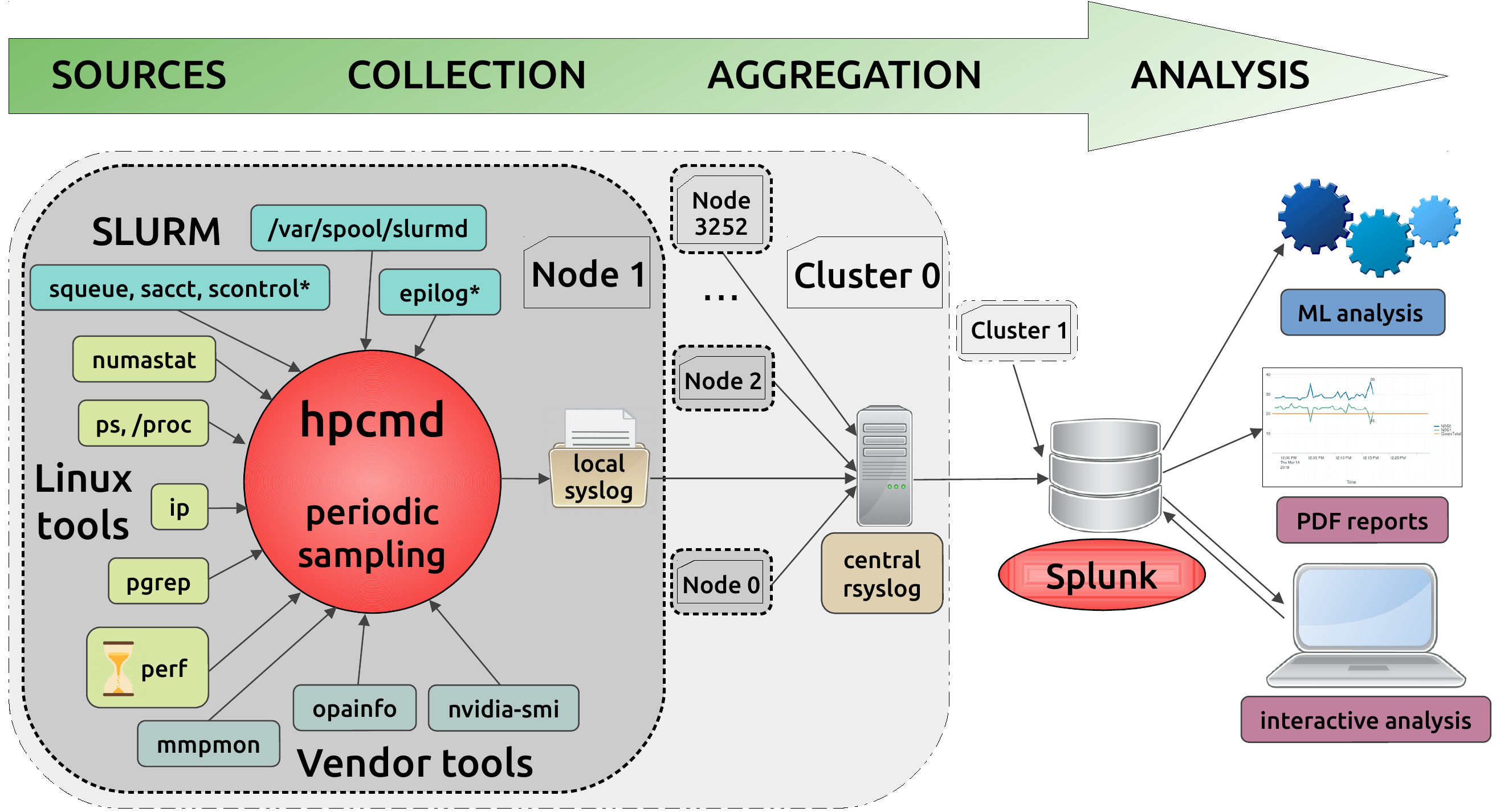}
\caption{Schematic showing the architecture of the MPCDF HPC monitoring system.
The \hpcmd middleware and various \splunk analysis dashboards were written by
the authors, while the other infrastructure had already been existing. Automatic
analysis using machine learning techniques is under development.}
\label{fig:architecture}
\end{figure}
Figure \ref{fig:architecture} presents a schematic overview on the
architecture of the MPCDF HPC monitoring system. The design was motivated by
the principle of simplicity and the focus on key questions which implied the
reuse of existing infrastructure.
To this end, our programming efforts focused on two major components (shown
with red background in Figure \ref{fig:architecture}).

The first component, labeled \hpcmd,
is a lightweight middleware that runs as a daemon in the background
on each compute node, performs measurements at regular intervals, and
computes derived metrics if necessary.
A thorough evaluation of the overhead of \hpcmd showed that the impact on the
application performance is negligible, e.g., being much smaller than the
influence of unavoidable machine and OS jitter on the application runtime.
\hpcmd is written in plain Python (both
versions 2.7 and 3 are supported) and configurable via a flexible,
hierarchical YAML configuration file.
Measured values are simply written by \hpcmd to syslog messages, forwarded via
rsyslog, and finally fed into a \splunk repository.

The second major component are dashboards for \splunk written in XML, that we
have developed for performance data analysis and visualization.
The \splunk \cite{SPLUNK} platform excels in the analysis of large volumes of
temporally ordered log-line data via a powerful query language. Hence,
\splunk is suitable for crunching performance data collected from many nodes
over long periods of time. After having collected performance data for nearly
one year now, we do not notice any performance degradation and do not see
any reason to limit the storage lifetime of the data.
There are several viable alternatives to \splunk which could be used
similarly, such as the open-source ELK (Elasticsearch, Logstash, Kibana) stack,
the InfluxDB-Grafana stack, or even custom frameworks.
However, the \splunk infrastructure was already installed and used at MPCDF
systems for system monitoring, and thus it was the natural choice to employ
it for performance monitoring as well.
We are considering the aforementioned alternative solutions for evaluation in the future.

In the rest of this section, we detail on the different aspects of the
data flow as depicted on the top of Figure \ref{fig:architecture}.

\subsection{Data Sources}
\label{sec:sources}

From a technical point of view, today's HPC hardware and software offer a
plethora of metrics to look at.  Given these possibilities,
it is necessary to carefully choose a set of observables essential to
yield valuable insight, and at the same time, keep the impact on the
application performance negligible and the data volume of the measured
values tractable.  \hpcmd uses the following data sources.

\paragraph{CPU core events:} State-of-the-art CPUs provide performance
monitoring units (PMUs) for each core. These can be programmed to
count events, e.g., scalar and vector instructions, cache
misses, and many more. Since the PMUs are additional programmable
hardware units, the event counting induces only minimal overhead,
typically not noticeable for the running scientific application.

\paragraph{CPU uncore events:} In addition to the core PMUs, modern CPUs
provide uncore PMUs that enable to monitor, e.g., the memory controller traffic
and the traffic between different sockets.
To access the core and uncore counters, the Linux \emph{perf} subsystem
is used.

\paragraph{GPU:}
At present, monitoring GPUs is much more difficult than CPUs due to
the dependencies on proprietary tools and APIs from hardware vendors,
as well as due to the lack of publicly available counter specifications.
Nevertheless, it is
possible to track some values, such as the memory occupation and the overall
utilization, which we do using the \emph{nvidia-smi} tool.

\paragraph{I/O:} Large parallel file systems are crucial
components of any HPC system. They are a shared resource, and wrong
usage may affect not only a single problematic job but potentially even
the whole system. Monitoring the I/O traffic and characteristics per
node can give valuable hints at harmful use patterns.
Since Spectrum Scale (GPFS) is the preferred file system at MPCDF, its CLI tools are used for monitoring. 

\paragraph{Network:} High-speed networks represent the backbone of
an HPC system. Communication characteristics at per-node resolution complement
many other metrics with valuable insight.  Relevant counters can be queried
using the CLI tools that come with InfiniBand or OmniPath network adapters.

\paragraph{Software:} The Linux kernel complemented by
various system tools gives access to a rich set of application-related
metrics, e.g., the number of tasks (processes and threads) actually launched
by the job, the pinning of these tasks, the memory usage, the job's environment
variables, and many more. The \hpcmd software accesses this kind of information using
the \emph{ps} and \emph{numastat} tools, and the \emph{/proc}
virtual filesystem in some cases.

Any of these observables can be sampled at regular intervals by \hpcmd and
used directly or after some arithmetic manipulation (e.g., computing the
GFLOP/s) as performance metrics.

\subsection{Data Collection}

On each compute node, an instance of the \hpcmd middleware is running in the
background as a systemd service, measuring at regular intervals and sequentially collecting data from the aforementioned sources.
The measurements are synchronized across the nodes via the system clock,
avoiding any communication between nodes.
In addition to a continuous operation mode, the \hpcmd daemon supports the
widely used SLURM batch system\cite{SLURM},
and determines the state of a node (allocated, idle, shared) and job information automatically.
We are typically monitoring only nodes which have a single job running on
them, i.e., data is not collected for nodes that are currently idle or
shared, as such cases are considered less relevant in our context and would be much
harder to interpret.
\hpcmd allows for a highly flexible configuration, e.g., to perform more
frequent sampling or per-core monitoring of performance counters.
Moreover, users may suspend the \hpcmd systemd service during the runtime of
a job to get exclusive access to hardware counters, e.g., for running performance
profilers such as VTUNE or using libraries such as PAPI.
Measured values and derived metrics are written as log lines containing
key-value pairs to the local syslog file.
For further details, we kindly refer the reader to the documentation of
\hpcmd\cite{HPCMDDOC}.

\subsection{Data Aggregation}

From each monitored node, the \hpcmd log lines are transported via rsyslog,
collected, and finally fed into a central \splunk system. At MPCDF, HPC
systems are configured such that the rsyslog traffic goes via the Ethernet
link, not putting any load on the high performance network reserved for the
applications. For large HPC systems, there may be intermediate (per-``island'')
rsyslog servers.
Operating at sampling intervals on the order of minutes we do not see any
scalability issues for our present and future HPC cluster sizes. See
Sec.~\ref{sec:studies} for some practical experience.

\subsection{Data Visualization and Interactive Analysis}

Data visualization and analysis takes place in the \splunk system, for which
we have developed several dashboards providing views at different levels of detail.

\paragraph{Roofline View:}
\begin{figure}[htb]
\centering
\includegraphics[width=\textwidth]{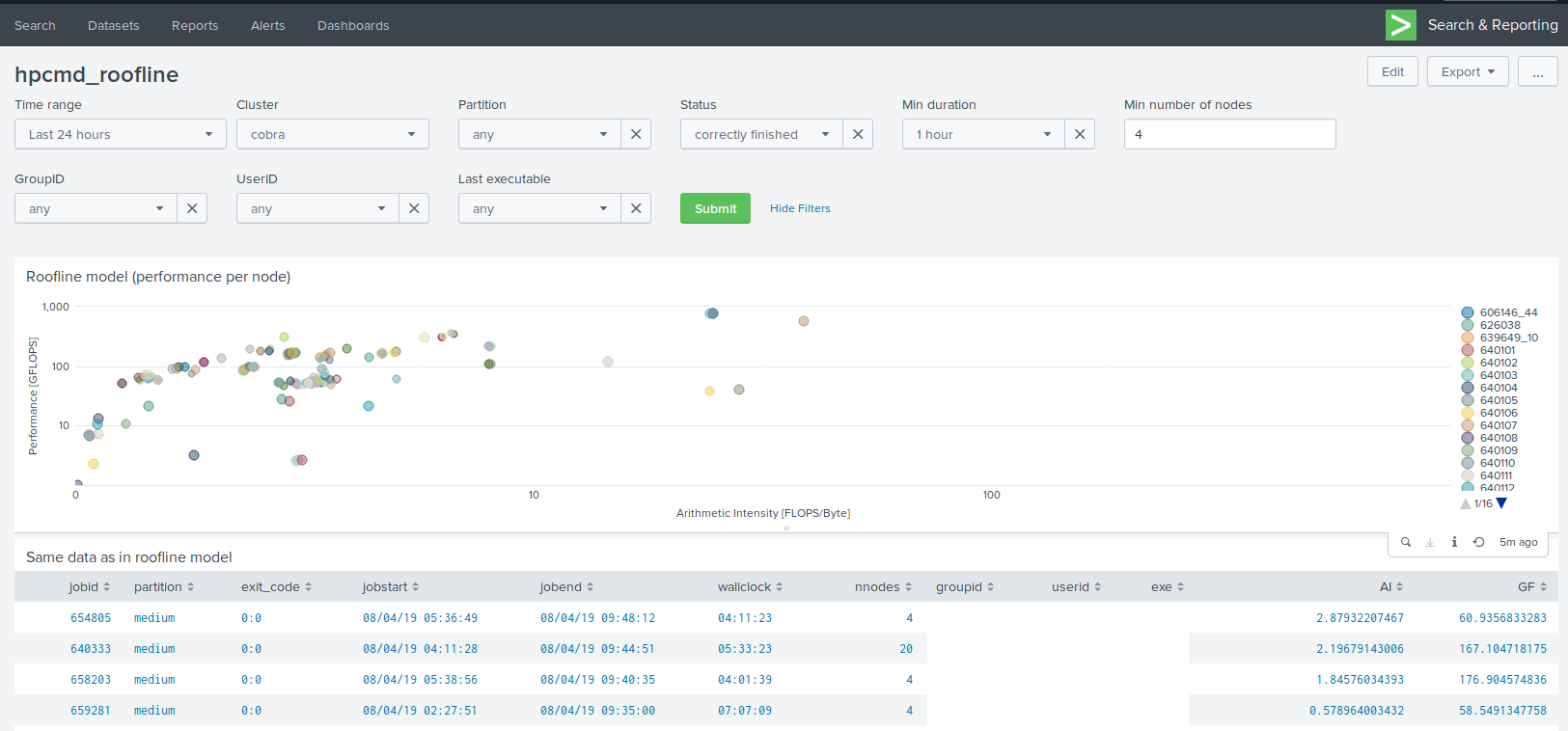}
\caption{Overview on a selection of jobs from the previous 24 hours in
a roofline plot on a specific HPC system. Each circle represents a
job with its average performance, where the circle sizes are scaled
by the actual CPU core hours of the jobs.}
\label{fig:roofline}
\end{figure}
The roofline model is a simple yet intuitive performance model widely
used in performance engineering \cite{ROOFLINE2009}.
This type of overview is suitable in particular when the performance of a job
needs to be condensed into only two numbers and related to the theoretical peak
values of the machine.
In a 2d system of coordinates, the horizontal axis denotes the arithmetic
intensity in FLOP/Byte, while the vertical axis denotes the performance in GFLOP/s.
%
We pragmatically chose to solely rely on CPU-RAM memory bandwidth for the roofline plot,
computed from CPU uncore events.
For the application support staff, the entry point for the inspection of
performance data in \splunk is a roofline-type of overview plot, as shown in
Figure \ref{fig:roofline}. All finished jobs that fall into a certain time
frame and satisfy certain constraints, which are specified by the user using drop-downs on the top of the
web-page, are displayed as colored circles, scaled in size by their
consumption of CPU hours.
This dashboard represents an intuitive performance map showing the current or
historic utilization status of the system. Clicking on a circle in the plot or on a line
in the data table below forwards to the detailed job view.

\paragraph{Detailed Job Views:}
\begin{figure}[htb]
\centering
\includegraphics[width=\textwidth]{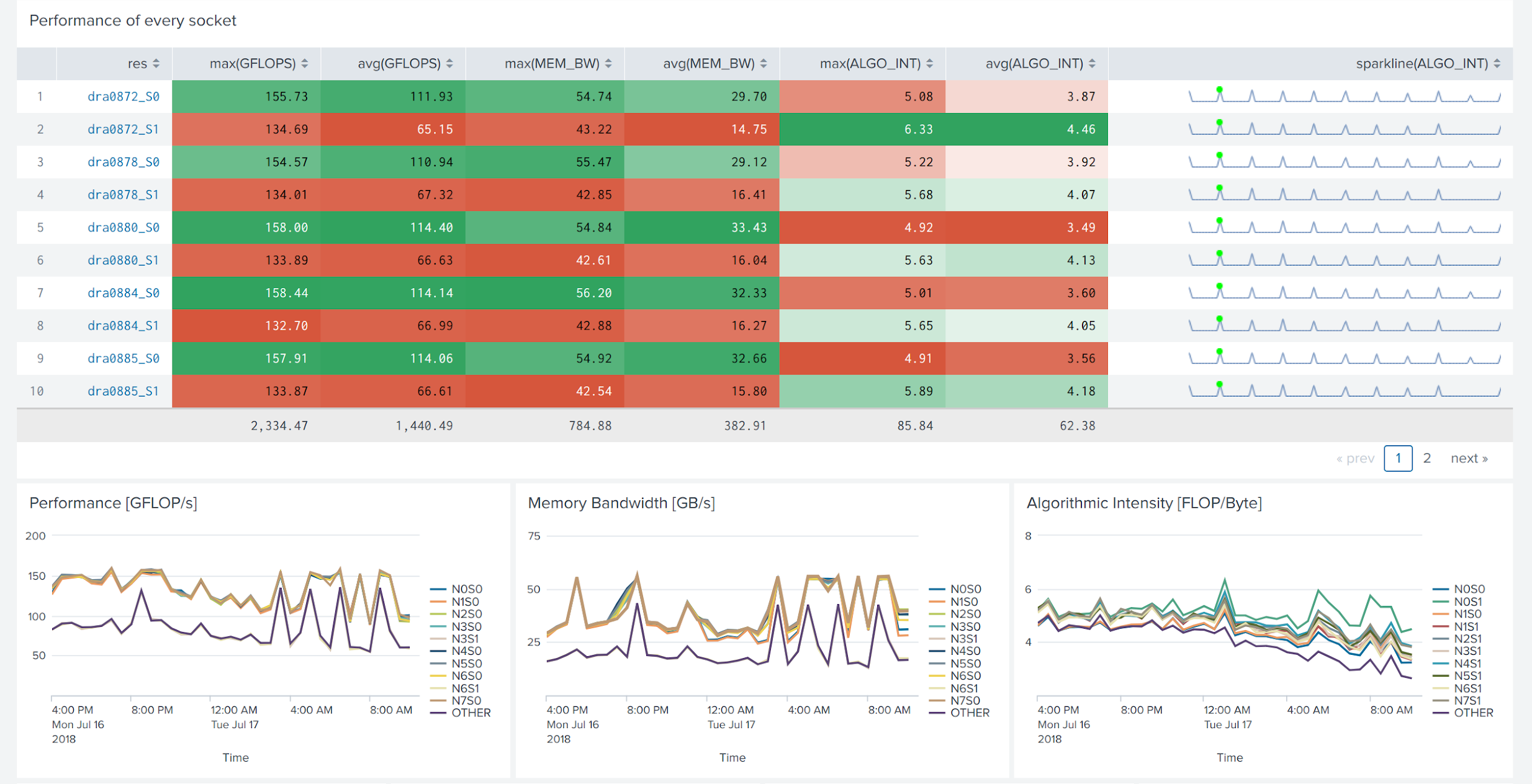}
\caption{Excerpt from a detailed view on a specific job, showing
the achieved performance in GFLOP/s, the memory bandwidth, and the
algorithmic intensity for each socket. In
addition to the averaged and maximum values shown in the table,
plots over time are available per socket.
Moreover, the \splunk dashboard contains about 30 more
plots for other CPU, GPU, network, filesystem, and software metrics
(not shown here).}
\label{fig:jobperf}
\end{figure}
This dashboard provides a detailed view on the job's performance
characteristics through temporal plots of the performance metrics described
in Subsection \ref{sec:sources}. An excerpt from the dashboard is shown in
Figure \ref{fig:jobperf}.
To make the data from large jobs more comprehensible, a second dashboard
is provided that displays the data using statistical variables such as
maximum, median, and minimum curves, taken from all nodes or sockets.
These two dashboards are intended to be used by the application support staff
through the interactive \splunk web interface.  For the users, static
PDF reports are provided for download containing the same information.
Based on these detailed job views it is typically possible to draw
well-grounded conclusions about performance issues of application codes.

\paragraph{Specialized Views:}
System administrators and the management of a computing center are
often interested in specific analysis of many jobs. To obtain such
information, they can submit custom queries to the \splunk database.
As some of their questions are recurrent,
we have developed several dashboards to ease their access to
the data. Currently we are providing plots that show the most
executed applications by core hours, jobs that reserved GPU nodes
without using GPUs, jobs that reserved large memory nodes without
using much memory, and jobs that use less than half of the available CPU cores.

\subsection{Per-job Reports for Users}
\label{sec:pdfreport}

To make the performance data accessible to the users, a performance report can be generated
for each job and
provided as a PDF file for download via a web server after login. We decided not to grant the
users access to \splunk directly for security, data protection, and administrative reasons.

\subsection{Data Analytics and Automation}

On the MPCDF HPC systems, several thousands of jobs are typically run per
day. To be able to cope with these numbers and the massive amount of
generated data, an automatic data analytics system is indispensable in
order to identify problematic jobs on the systems, and notify both
support staff and users in critical cases. The data analytics module
of the HPC monitoring system is currently under development, but goes
beyond the scope of this paper.

\section{Scenario- and Case-Studies}
\label{sec:studies}

The HPC monitoring system is used to continuously monitor the HPC systems DRACO
($\approx$ 940 nodes, $\approx$ 32K cores) and COBRA ($\approx$ 3250 nodes,
$\approx$ 130K cores) at the MPCDF.
These HPC systems are heterogeneous, containing nodes with different CPU
micro-architectures, with different RAM sizes, and with or without GPU
accelerators of different models.
The system is configured to write performance data every 10 minutes which
generates up to 3 KiB of raw log line data per node. Hence, the total
data volume per sample for both machines is about 12.5 MiB, which amounts
to about 1.8 GiB per day in total. Note that the rsyslog system is able to
easily cope with that data volume, making complex custom hierarchical transport
agents unnecessary in our case.
In the following, we illustrate with 4 examples how the HPC monitoring system already proved to be
helpful in practice at the MPCDF.

\paragraph{Suboptimal Job Scripts:}
We provide users with a detailed job-specific report (see Subsection
\ref{sec:pdfreport} for more details), based on which they can quickly spot
potential errors related to their job scripts. We are aware of several cases
where HPC monitoring was already helpful in this respect.

\paragraph{Hanging Jobs:}
Even though HPC clusters are supposed to be used to run
stable programs, there are still jobs that encounter problems at
runtime without shutting down in a controlled manner.
For example, in cases of livelocks or deadlocks, the processes of a
job continue to run without actually executing any useful
instructions, thereby occupying the reserved resources. This can
potentially waste a large number of CPU hours.
Such ``hanging'' jobs are typically manifested
by very low values in certain performance metrics, especially in
GFLOP/s and IPC.
To report on a specific example, it was observed from the HPC
monitoring data that jobs from a particular user
often demonstrated the aforementioned behavior. We contacted
the user and showed the plots that illustrated the performance
problem. The user then investigated the code and fixed the issue.
Catching this particular case was achieved unintentionally, by manual inspection of the
data, however an automatic detection system for such types of jobs is
under development.

\paragraph{Verification of the Utilization of Extra Resources:}
To satisfy the compute needs for a broad spectrum of users, computing centers
often equip parts of their HPC systems with nodes that contain very large
amounts of RAM memory or with nodes that contain GPU accelerators.
Sometimes, users with applications that require only moderate amounts of
memory or lack GPU support, by mistake or by convenience, allocate such nodes
with extra resources instead of regular ones.
This is not a problem if these nodes would otherwise be idle, but if not, such
allocations mean a waste of resources and increased queueing times for
legitimate users.
HPC monitoring can easily detect this type of wrong usage and warn staff or the users
directly.

\paragraph{Coarse-grain Overview for Experts:}
The HPC monitoring system has not been designed for in-depth code
profiling. Nevertheless, it can still provide coarse-grain performance
information that can be useful to code developers and
application support.
Indeed, several members of the application support group at the MPCDF routinely use
HPC monitoring to inspect the performance of
applications they personally contributed to during development. In most
cases, HPC monitoring confirmed their expectations.
Interestingly, there were some occasions when even these experts were
surprised. In fact, the \splunk analysis of the data showed that the
performance in some stages of the
application was much worse than expected, which had notable
influence on the overall runtimes of the programs. The reason was
the lack of code vectorization for some code
blocks that were initially considered less relevant.
As a next step, the developers profiled the code with more specialized tools
which confirmed the observation from the \splunk dashboards and were able to
point to the exact lines of code that caused the performance issue.

\section{Summary and Outlook}
\label{sec:summary}

This paper reports on the requirement analysis, the design, and the
implementation of the MPCDF HPC performance monitoring system.
Our solution is simple, modular, lightweight,
mostly based on standard Linux tools, and thus it can easily be
adopted by other HPC centers.
The system is
in operation to comprehensively monitor the performance of all jobs running
on two large HPC systems at the MPCDF with about 4200 nodes and more than
160.000 CPU cores in total.
After several months of production we have collected a large
amount of job-related performance data, and doing
data analytics on it will be the main topic of our future work.
Additionally, we plan to extend the deployment of our performance monitoring
system to more (medium-sized) clusters at the MPCDF, and will
continue to develop and maintain the \hpcmd middleware.

\paragraph{Acknowledgments:}
We are grateful to Christof Hanke for the continuous support with
\splunk and the implementation of major parts of the PDF generation web
service.
We are indebted to Alexis Huxley and Christian Guggenberger for the
regular (re)installation of the HPC monitoring software on the HPC
systems at short notice.
Finally, we thank Lorenz H\"udepohl, Andreas Marek, Pavel Kus,
Sebastian Ohlmann, and Markus Rampp from the application support group
for many valuable suggestions and fruitful discussions.
The final authenticated publication is available online at
\url{https://doi.org/...}.

\paragraph{Software:}
The \hpcmd software is free of charge and publicly available for download at
\codeurl. Online documentation is available at \docurl. The software is
licensed under the permissive MIT license. We kindly request to cite this
paper in case the software is used and reported on in publications.

\bibliographystyle{splncs04}
\bibliography{hpcmd}

\end{document}